\documentclass[11pt,a4paper]{article}
\usepackage{jheppub}
\pdfoutput=1
\usepackage{multirow}
\usepackage{makecell}
\usepackage{graphicx}

\subheader{UTTG-17-16}
\title{Ordered Kinematic Endpoints for 5-body Cascade Decays}
\author[a]{Matthew D. Klimek}
\affiliation[a]{Theory Group, Department of Physics and Texas Cosmology Center,\\
University of Texas at Austin,\\
Austin, TX 78712, USA}
\emailAdd{klimek@physics.utexas.edu}

\abstract{
	We present expressions for the kinematic endpoints of 5-body cascade decay chains proceeding through all possible combinations of 2-body and 3-body decays, with one stable invisible particle in the final decay stage.  
	When an invariant mass can be formed in multiple ways by choosing different  final state particles from a common vertex, we introduce techniques for finding the sub-leading endpoints for all indistinguishable versions of the invariant mass.
	In contrast to short decay chains, where sub-leading endpoints are linearly related to the leading endpoints, we find that in 5-body decays, they provide additional independent constraints on the mass spectrum.
}
\notoc
\begin{document}
\maketitle
\flushbottom

\newcommand{\mw}{m_W}
\newcommand{\mx}{m_X}
\newcommand{\mxtil}{m_{\tilde X}}
\newcommand{\mxtils}{m_{\tilde X}^2}
\newcommand{\mztil}{m_{\tilde Z}}
\newcommand{\mztils}{m_{\tilde Z}^2}
\newcommand{\my}{m_Y}
\newcommand{\mz}{m_Z}
\newcommand{\ma}{m_A}
\newcommand{\mas}{m_A^2}
\newcommand{\mb}{m_B}
\newcommand{\mbs}{m_B^2}
\newcommand{\mc}{m_C}
\newcommand{\mcs}{m_C^2}
\newcommand{\mxs}{m_X^2}
\newcommand{\mys}{m_Y^2}
\newcommand{\mzs}{m_Z^2}
\newcommand{\mws}{m_W^2}
\newcommand{\partX}{ X}
\newcommand{\partY}{ Y}
\newcommand{\partZ}{ Z}
\newcommand{\partA}{ A}
\newcommand{\partB}{ B}
\newcommand{\partW}{ W}
\newcommand{\sqx}{\mx}
\newcommand{\sqy}{\my}
\newcommand{\sqz}{\mz}
\newcommand{\high}{\mathrm{high}}
\newcommand{\highest}{\mathrm{highest}}
\newcommand{\low}{\mathrm{low}}
\newcommand{\lowest}{\mathrm{lowest}}
\newcommand{\mymz}{\frac{\my}{\mz}}
\newcommand{\mxmy}{\frac{\mx}{\my}}
\newcommand{\mwmx}{\frac{\mw}{\mx}}
\newcommand{\mymzs}{\frac{\mys}{\mzs}}
\newcommand{\mxmys}{\frac{\mxs}{\mys}}
\newcommand{\mwmxs}{\frac{\mws}{\mxs}}

\section{Introduction}
The measurement of kinematic endpoints is a classic technique for determining the masses of new particles in collider contexts \cite{Barr:2010zj,Gjelsten:2004ki,Allanach:2000kt,Matchev:2009iw,Burns:2009zi,Gjelsten:2005aw}.  It is applicable to situations in which a heavy parent particle decays, possibly through one or more mediators, to visible Standard Model particles and one or more stable invisible states.  Such decays are a generic prediction of many models of physics beyond the Standard Model (R-parity conserving SUSY \cite{Gjelsten:2004ki,Allanach:2000kt,Matchev:2009iw,Burns:2009zi,Gjelsten:2005aw}, T-parity in little Higgs models \cite{Cheng:2003ju,Perelstein:2005ka}, extra dimensions \cite{Appelquist:2000nn,Hooper:2007qk}, etc.).  The location of the endpoints depends only on the mass spectrum of the involved particles and is independent of their other properties, making the technique especially suitable for collider discoveries where spin, CP properties, and couplings may not be immediately known.

The endpoints for decays to four final state particles have been studied extensively \cite{Gjelsten:2004ki}, often motivated by scenarios involving squark production followed by decays to one quark, two leptons, and an invisible LSP, or other scenarios with equivalent final states.  Also of interest are scenarios such as gluino production which can give cascade decays with five final state particles $\tilde g\to qqll+\mathrm{LSP}$, or others with equivalent final states. The endpoints for the 5-body decay consisting of a chain of sequential 2-body decays where all intermediate particles are on-shell have been presented by \cite{Gjelsten:2005aw}.   

In general, some of the mediators may be off-shell.  When integrated out, these can be considered as vertices from which more than one visible particles are emitted. These particles are kinematically indistinguishable, even if they are different species.\footnote{In practice, particles such as jets may be indistinguishable in a collider context even when they are not from the same vertex, but this is not an issue intrinsic to the kinematics and will not be treated here.}
This gives rise to multiple ways of forming invariant masses involving particles from that vertex. While these invariant masses will be indistinguishable, they can be ordered.  
Consider computing all possible indistinguishable versions of one of these invariant masses and inspecting the distribution of the $n$-th largest version in each event.  The endpoint of this distribution will be the $n$-th ordered endpoint.   

\begin{figure}
\includegraphics[scale=0.65]{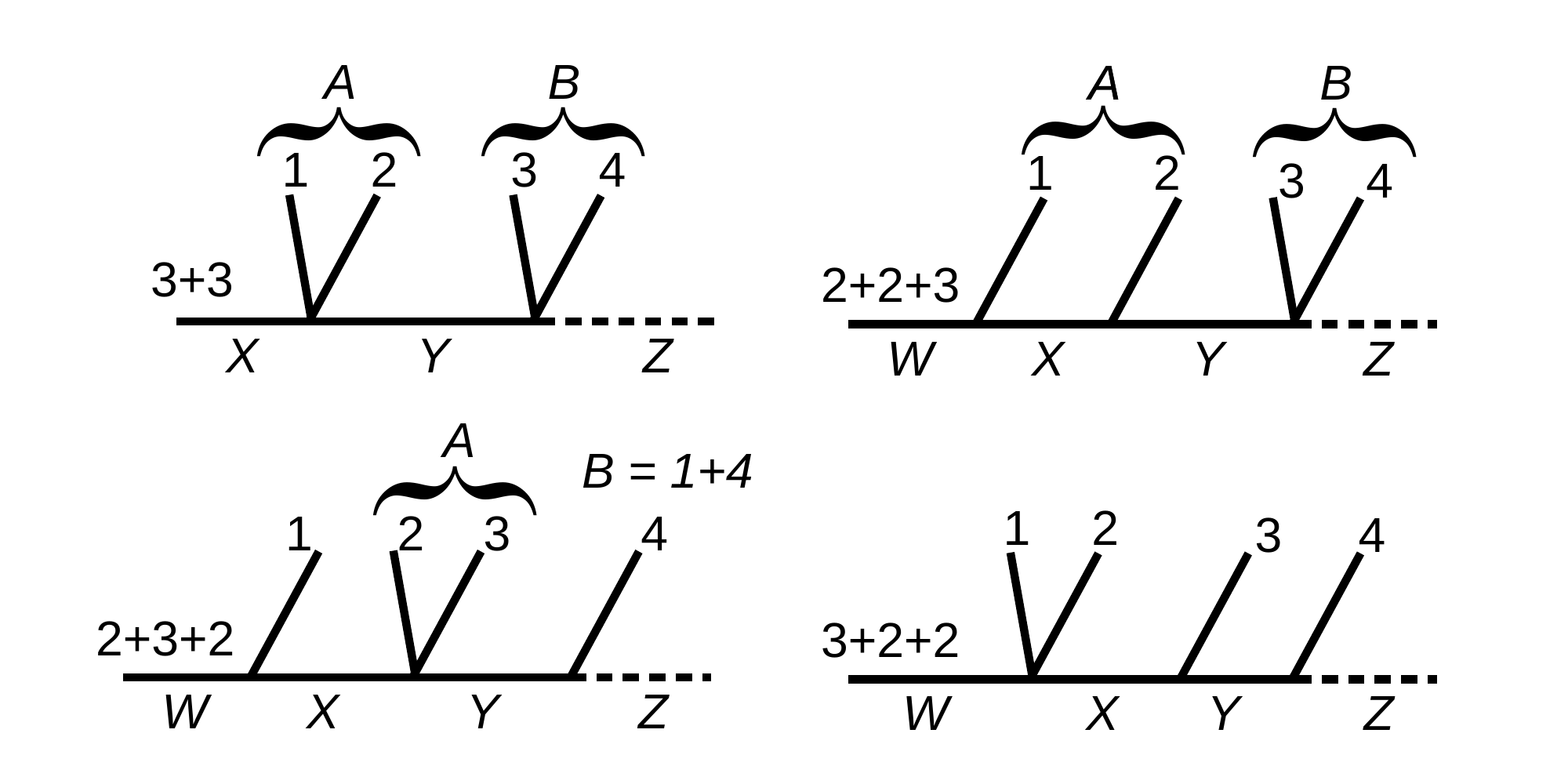}
\caption{All 5-body decays with one or more 3-body vertices.  The labels A and B are used throughout the text to refer to groupings of particles when convenient.}\label{decaysfig}
\end{figure}

The ordered endpoints for direct $1\to 5$ decay, where all particles are emitted from a common vertex, have already been discussed in \cite{Kim:2015bnd}.  Here, we consider decays which proceed through one or more intermediate 3-body decays with the rest of the chain being composed of 2-body decays\footnote{Although in principle an intermediate 4-body decay could appear, the width for such decays is highly suppressed by the masses of the off-shell mediators, so that the branching ratio will typically be small relative to other decay modes that allow for on-shell mediators.}
and with one invisible particle at the last stage.  
These are depicted in figure \ref{decaysfig}.

In order to solve for a number of unknown masses, it is necessary to measure at least that many linearly independent endpoints, but it is preferable if possible to measure more. The resulting over-constrained system provides a check on assumptions about the decay topology.  The precision with which endpoints can be measured over background with small signal samples is limited by the fact that the endpoints of many invariant mass distributions lie at the end of long tails.  Measuring more linearly independent endpoints than strictly necessary also helps to improve uncertainties in these cases.
In 4-body decays including one 3-body vertex, the ordered endpoints of sets of indistinguishable invariant masses are related to each other by a constant factor.  As such, the sub-leading endpoints do not provide any additional information that can be of use in constraining the masses present in the decay.
However, for the 5-body decays that we will consider here, the expressions for the sub-leading endpoints generally are not linearly related to those of the leading endpoints. They thus provide valuable additional information that would otherwise be lost if they were ignored.  

This paper is organized as follows. In \S\ref{conventions} we list various conventions that will be followed in the calculations we present. In \S\ref{33sec} we present the calculations for endpoints in the 3+3 decay chain.  In \S\ref{223sec} we present the endpoint calculations for the other three decay chains containing three vertices.  We conclude in \S\ref{conclusion}.  Appendices follow containing further details of some of the calculations.


\section{Conventions}\label{conventions}
We adopt a convention of labelling the different decay chains by the multiplicity of each sequential vertex, so that, for example, the decay consisting of two 2-body decays followed by a final 3-body decay is called 2+2+3.


It will often be convenient for our calculations to combine two particles into a single composite object, or ``pseudoparticle.''  For example, combining two particles with 4-momenta $p_1^\mu$ and $p_2^\mu$, we obtain a composite particle which we denote as $\partA=1+2$ with 4-momentum $p_A^\mu=p_1^\mu +p_2^\mu$.  We could also ``subtract'' two particles giving a composite particle $\partA=1-2$ with 4-momentum $p_A^\mu=p_1^\mu-p_2^\mu$.

Throughout the text, lower case letters such as $p$ and $k$ without a Lorentz index stand for the magnitude of the 3-momentum of the corresponding object, $p\equiv |{\bold p}|$, etc.  We assume that all visible stable particles are massless.

We will make extensive use of the kinematic function
\begin{equation}\label{kinematicfn}
	[A,B,C]=\frac{ \ma^4+\mb^4+\mc^4-2(\mas\mbs+\mbs\mcs+\mas\mcs) }{4\mbs}
		= \frac{(\mas-\mbs-\mcs)^2 - 4\mbs\mcs}{4\mbs}
\end{equation}
which gives the squared magnitude of the 3-momentum of any two particles with mass $\ma$ and $\mc$ involved in a 2-body decay in the rest frame of the third particle with  mass $\mb$:
\begin{equation}
	p_A^2=p_C^2=[A,B,C],\quad p_B^2=0.
\end{equation}
The energy of one of these particles is then
\begin{equation}
	E_A^2 = p_A^2 + \mas = [A,B,C]+\mas = \frac{(\mas+\mbs-\mcs)^2}{4\mbs} .
\end{equation}

Endpoint studies are often motivated by SUSY scenarios whose phenomenology is characterized by gluino or squark decay chains terminating in an LSP.  For this reason, endpoints such as those considered here often appear in the literature with SUSY-based naming conventions \cite{Gjelsten:2004ki,Gjelsten:2005aw}.  In this work, we use a more general labeling convention, but for comparison with other works cited here, the following translation can be used:
\begin{equation}
	1\to q_\mathrm{n},\quad
	2\to q_\mathrm{f},\quad
	3\to l_\mathrm{n},\quad
	4\to l_\mathrm{f}.\quad
\end{equation}

\section{The 3+3 decay chain}\label{33sec}
First we will consider the 3+3 decay chain depicted in the first panel of figure \ref{decaysfig}. The decay begins with a particle $\partX$ with mass $\mx$ which decays to two massless particles, 1 and 2, and a massive particle $\partY$ with mass $\my$, which in turn decays to two more massless particles, 3 and 4, and a stable particle $\partZ$ with mass $\mz$ assumed to be invisible.  When convenient, we will  consider particles 1 and 2 together as a composite object which we will denote $\partA=1+2$, and likewise for particles 3 and 4 which we denote $\partB=3+4$.  All calculations will be performed in the rest frame of the $\partY$ particle.

There are 11 kinematic invariants that can be formed with four visible final state particles: 6 pairs ($m_{12}^2$, $m_{13}^2$, $m_{14}^2$, $m_{23}^2$, $m_{24}^2$, and $m_{34}^2$), 4 triplets ($m_{123}^2$, $m_{124}^2$, $m_{134}^2$, and $m_{234}^2$), and 1 quadruplet ($m_{1234}^2$).  Four of the pairs ($m_{13}^2$, $m_{14}^2$, $m_{23}^2$, and $m_{24}^2$) are equivalent up to a choice of which particle is being selected from each stage of the decay.  We refer to these four collectively as $m_{13}^2$ and compute their ordered endpoints.
Similar considerations apply for the triplets which come in two indistinguishable pairs ($m_{123}^2$ and $m_{124}^2$; $m_{134}^2$ and $m_{234}^2$).

The results of this section will be collected in table \ref{33table}.

\subsection{Endpoints for pairs $m_{12}^2$ and $m_{34}^2$}
Considering particles 1 and 2 together as the composite particle $\partA=1+2$, it is clear that $m_{12}^2=\mas$ cannot exceed the value 
\begin{equation}
	{
		\max(m_{12}^2)=(\sqx-\sqy)^2.
	}
\end{equation}
This is achieved when particles 1 and 2 are back to back with equal momenta $p_1=p_2=(\mx-\my)/2$ and particles $X$ and $Y$ are both at rest.
Similarly, 
\begin{equation}
	{
	\max(m_{34}^2)=(\sqy-\sqz)^2.
}
\end{equation}

\subsection{Ordered endpoints for pairs equivalent to $m_{13}^2$}\label{33m13}
As discussed above, we can form four equivalent invariant masses of this type by choosing one visible  particle from each of the two vertices. We will call the ordered endpoints $\max(m_{13}^2)_\highest$, $\max(m_{13}^2)_\high$, $\max(m_{13}^2)_\low$, and $\max(m_{13}^2)_\lowest$.

It can be shown that the $n$-th ordered endpoint will occur when  $n$ of the indistinguishable endpoints are equal.  
For example, in the case of $\max(m_{13}^2)_\high$, we may write
\begin{equation}
	(p_X^\mu - p_Z^\mu)^2 = \left(\sum_{i=1}^4 p_i^\mu\right)^2 = \sum_{i<j\le4} m_{ij}^2 ,
\end{equation}
or equivalently, in the rest frame of $X$,
\begin{equation}\label{pfeqn}
	\mxs+\mzs - 2\mx E_Z - m_{12}^2 - m_{34}^2 - m_{14}^2 - m_{24}^2 = m_{13}^2 + m_{23}^2 . 
\end{equation}
This relation depends on six kinematic degrees of freedom.  Fixing the five quantities on the lefthand side determines the value of the righthand side.   The sixth degree of freedom consists of choosing the partition of that value between $m_{13}^2$ and $m_{23}^2$.
The phase space for the 3+3 decay is 7-dimensional\footnote{For any 5-body decay, there are $5\times 3$ components of momentum.  Momentum conservation fixes 3, and the on-shell conditions for the parent and the mediator fix 2 more.  Finally, 3 rigid rotations are irrelevant.  This counting yields $15-3-2-3=7$.}  so these choices do not over-constrain the system.
To find the 2nd ordered endpoint, we must maximize $m_{13}^2$ without exceeding $m_{23}^2$. For any fixed value of the lefthand side, this clearly results in $m_{13}^2=m_{23}^2$.  This value of $m_{13}^2$ can then be maximized over all remaining degrees of freedom.
By moving $n$ invariant masses to the righthand side of the equation, we can repeat this argument for the $n$-th ordered endpoint.

\subsubsection{Highest endpoint}\label{m23highesttxt}
We will begin by finding the absolute maximum of the pair invariant mass.  Assume we have chosen to form the pair with particles 1 and 3.  We would expect the invariant mass of the pair to be maximized in the limit when both 1 and 3 take all of the available energy in their respective decays while their partners, particles 2 and 4, become soft $p_{2,4}\to 0$.  In this case, the decay chain reduces to an effective 3-body decay, with the well-known endpoint
\begin{equation}\label{33m13highestresult}
	{
		\max(m_{13}^2)_\highest = \frac{(\mxs-\mys)(\mys-\mzs)}{\mys},
	}
\end{equation}
obtained when particles 1 and 3 are aligned back to back.
This result is obtained by explicit computation in Appendix \ref{m23highestapp}.
We note in passing that the requirement of taking the soft limit of unused particles that share a vertex with the particles that compose the invariant mass whose highest endpoint we are finding is a general pattern which will hold for all similar endpoints.

\subsubsection{High endpoint}
As shown above, the highest possible value that one of the equivalent invariant masses can have without exceeding another must occur when the two highest are equal.  The righthand side of eq.~(\ref{pfeqn}) will be maximized by taking particles 1 and 2 to be collinear and particle 4 to have vanishing momentum.  In that case, all invariant masses on the lefthand side of eq.~(\ref{pfeqn}) are zero.
 In this configuration, the $A=1+2$ system is still massless and each constituent has half the momentum that particle 1 had in the previous case $p_1=p_2=\sqrt{[0,Y,X]}/2$.  As a result, we have 
\begin{equation}
	{
		\max(m_{13}^2)_\high =\frac{ \max(m_{13}^2)_\highest}{2}.
	}
\end{equation}
	  An equivalent result would be obtained if we instead considered the configuration where $p_3=p_4$ and they are recoiling against particle 1 or 2.

\subsubsection{Low endpoint}\label{m23lowtxt}
\begin{figure}
\includegraphics[scale=0.75]{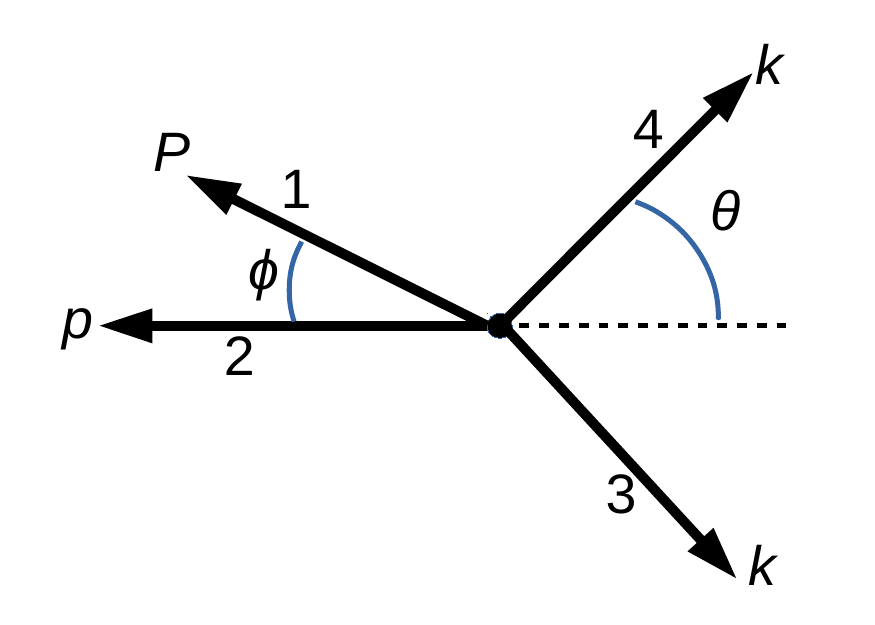}
\caption{Configuration for maximization of $(m_{13}^2)_\low$.}\label{m13lowfig}
\end{figure}

For the third lowest endpoint we want to find a configuration where three of the invariants are equal.  Symmetrical configurations of four particles result in either two or four invariants being equal.  A more complicated configuration is needed to give three equal invariants.  We can ensure that $m_{23}^2=m_{24}^2$  by configuring particles 3 and 4 to have equal momenta $k$ and particle 2 to have momentum $p$ and to make the same angle $\theta$ with respect to both 2 and 3.  Particle 1 can then be placed with momentum $P$ at some angle $\phi$ away from 2 in the same plane, as shown in figure \ref{m13lowfig}.\footnote{Particles 1 and 2 could be rotated so that particle 1 is not in the same plane as the others without changing the magnitude of their momenta.  However, due to the dot product in the definition of the invariant mass, it is always optimal to have as a large an angle as possible between the particles whose invariant mass we are trying to maximize.  This favors co-planar configurations.}
$P$ must be fixed as a function of $\phi$ so that $m_{13}^2$ will be equal to $m_{23}^2$ and $m_{24}^2$.  Particle 1 will make a smaller with particle 4 than with particle 3, so that $m_{14}^2$ will be less than the others.  We will denote the cosines of these angles as $b\equiv\cos\phi$ and $c\equiv\cos\theta$.

Energy and momentum conservation yield the requirements
\begin{equation}\label{firstloweq}
	\my = 2k+\sqrt{\mzs+(2kc)^2}
\end{equation}
and 
\begin{equation}
	p+P+ \my = \sqrt{\mxs+p^2+P^2+2pPb}.
\end{equation}
Requiring $m_{13}^2=m_{23}^2$ yields
\begin{equation}\label{lastloweq}
	p(1+c) = P \left(1+cb\sqrt{1-c^2}\sqrt{1-b^2} \right).
\end{equation}
This system of equations can be solved for $p$, $P$, and $k$, and the result substituted into the expression for the invariant mass
\begin{equation}
	m_{23}^2 = 2pk(1+c). 
\end{equation}
  The result is shown in Appendix \ref{m23lowapp}.  We must then maximize over $b$ and $c$.  Unfortunately, this must be done numerically.
For some values of the masses, a maximum is found in the region $0<(b,c)<1$, while for others, the maximum is on the boundary at $b=c=1$.  In this latter case, the low endpoint and the lowest endpoint coincide, as discussed below.

\subsubsection{Lowest endpoint}
The highest possible value of one of the equivalent invariant masses that does not exceed any of the others is obtained when all four are equal.  In this case the particles in each pair (1,~2) and (3,~4) are collinear and have equal momenta, and the two pairs are recoiling against each other.  It is easy to see that the result is 
\begin{equation}
	{
		\max(m_{13}^2)_\lowest = \frac{\max(m_{13}^2)_\highest}{4}.
	}
\end{equation}

\subsection{Endpoint for quadruplet $m_{1234}^2$}
The greatest amount of energy that could possibly be deposited into the (1234) system is $\sqx-\sqz$, which would occur when $\partX$, $\partY$, and $\partZ$ are all at rest in some frame.  In terms of the composite particles $\partA$ and $\partB$, we can write 
\begin{equation}
	m_{1234}^2= \mas+\mbs+2(E_A E_B-p_A p_B \cos\theta_{AB}).
\end{equation}
  The maximum values of $\ma$ and $\mb$ are $\sqx-\sqy$ and $\sqy-\sqz$, respectively.  When these values occur, we have have $p_A=p_B=0$ and $E_{A}=\ma$, $E_B=\mb$.  By explicit evaluation, we have for this case 
\begin{equation}
	{
		\max(m_{1234}^2)=(\sqx-\sqz)^2.
	}
\end{equation}
This form for the endpoint is valid for all mass spectra.  In contrast, the corresponding endpoint in other decay chains can have many different forms depending on the masses \cite{Gjelsten:2005aw}.  The simple behavior in this case is due to the existence of a pair of particles at each vertex which can always be arranged so that each pair has zero total momentum, allowing all massive particles in the decay to remain at rest.

\subsection{Ordered endpoints for triplets equivalent to $m_{123}^2$}
The analysis of the triplet invariant masses turns out to be fairly complicated, involving different solutions in different regions of parameter space.  We will also need to find the endpoint of the distribution of the lower of the equivalent triplet invariant masses in each event, in addition to the absolute maximum.  We will call these $\max(m_{123}^2)_\high$ and $\max(m_{123}^2)_\low$.

\subsubsection{High endpoint}\label{m123hightxt}
The calculation of this endpoint is facilitated by considering the composite particle $\partA=1+2$.  It is then equivalent to a two-particle endpoint $m_{123}^2=m_{A3}^2$ where the mass of $\partA$ is a free parameter in the range $\ma\in[0, \sqx-\sqy]$. 
We can combine the unused final state particle 4 with $\partZ$ to obtain a pseudoparticle $\tilde\partZ$, but for the same reason discussed in \S\ref{m23highesttxt}, we will want $p_4=0$ so that $\mztil=\mz$.  We will want particle 3 to be back to back with $\partA$ in order to maximize the invariant mass.  After accounting for this, we have $m_{123}^2=\mas+2p_3(E_A+p_A)$ where $E_A=\sqrt{\mas+p_A^2}$, $p_A^2=[X,Y,A]$, and $p_3^2=[Z,Y,0]$.  Using these definitions, we obtain  
\begin{equation}\label{33m123endpointexpn}
	m_{123}^2 = m_{A3}^2 = \mas+\frac{\mys-\mzs}{2\mys}\left(\mxs-\mys-\mas+\sqrt{(\mxs+\mys-\mas)^2-4\mxs\mys}\right).
\end{equation}
Maximizing this expression over $\mas$ gives 
\begin{equation}
	m_{123}^2=(\sqx-\sqz)^2,
\end{equation} 
which is the highest possible value for any invariant mass in this decay.  This occurs when 
\begin{equation}\label{m123mavalue}
	\mas = \mxs+\mys - \frac{\mx}{\mz}(\mys+\mzs).
\end{equation}
If $\mx/\my < \my/\mz$ this implies that the maximum occurs at unphysical values $\mas < 0$.  In this case, the maximum remains on the $\mas=0$ end of the physically allowed range and gives 
\begin{equation}\label{m123altendpoint}
	m_{123}^2=\frac{(\mxs-\mys)(\mys-\mzs)}{\mys}.
\end{equation}

These results are easy to understand physically.  When $\mas$ can take the value given by eq.~(\ref{m123mavalue}), it is able to exactly balance the momentum of particle 3, so that $X$ and $Z$ are both at rest in the same frame.  This gives the maximum possible invariant mass. (Note that $Y$ is not at rest in this frame. Rather, particle 3 carries away all of its momentum so that $Z$ emerges at rest from the $Y$ decay.) As $\mx/\my$ decreases relative to $\my/\mz$, eventually there is insufficient energy available for $A$ to balance particle 3.  The best that can be done is for particles 1 and 2 to be collinear and back to back with particle 3, making $A$ massless, and reducing the problem to a 3-body decay with the familiar endpoint eq.~(\ref{m123altendpoint}).


\subsubsection{Low endpoint}\label{m123lowtxt}
The low endpoint has a rich structure, but can be expressed in relatively simple analytical formulae.
For reasons identical to those given in \S\ref{33m13}, it will correspond to a configuration where $m_{123}^2=m_{124}^2$.  
Indeed, eq.~(\ref{pfeqn}) can be modified to read
\begin{equation}
	\mxs+\mzs - 2\mx E_Z + m_{12}^2 - m_{34}^2 = m_{123}^2 + m_{124}^2 , 
\end{equation}
from which the argument proceeds as before.
Although we are interested in particles 3 and 4 separately, it is convenient to express the intermediate formulae in terms of the composite $\partB=3+4$. 

In Appendix \ref{m123lowapp}, eq.~(\ref{33m123lowendpointexpn}), we find $m_{123}^2$ as a function of the composite particle  masses $\ma$ and $\mb$ subject to the constraint $m_{123}^2=m_{124}^2$.
The range of physical values of $\ma$ and $\mb$ is a rectangle $\ma\in [0,\sqx-\sqy]$, $\mb\in [0,\sqy-\sqz]$.
It remains to maximize the invariant mass over this region. The maximum always occurs on the boundary.  There are two candidate extrema.  

One occurs in the corner $(\ma,\mb)=(\sqx-\sqy,\sqy-\sqz)$ and gives
\begin{equation}\label{33m123extremA}
	m_{123}^2 = (\sqx-\sqy)(\sqx-\sqz) .
\end{equation}
This corresponds to a configuration in which each pair of visible particles is back to back with equal momenta in the $Y$ frame and with an arbitrary angle between the direction of the two pairs, or equivalently, where $A$ and $B$ as well as $X$, $Y$, and $Z$ are at rest.  Note that this is the configuration which gives the maximum possible value $(\mx-\mz)^2$ to invariant masses such as $m_{1234}^2$.  The value of eq.~(\ref{33m123extremA}) is lower since one particle is being left out of this invariant mass.

The other occurs at the value of $\ma$ given in eq.~(\ref{33m123edgemavalue}) along the edge $\mb=0$ and gives
\begin{equation}\label{33m123extremB1}
	m_{123}^2 = \left(\sqx - \sqrt{\frac{\mys+\mzs}{2}}\right)^2 .
\end{equation}
This corresponds to the same configuration found in \S\ref{m123hightxt}, but with particles 3 and 4 collinear and sharing the momentum equally.
The value of $\mas$ in eq.~(\ref{33m123edgemavalue}) becomes negative if
\begin{equation}
	\mxmys < \frac{2\mys}{\mys+\mzs}.
\end{equation}
 and the extremum saturates to the point $(\ma,\mb)=(0,0)$ with a value
\begin{equation}\label{33m123extremB2}
	m_{123}^2 = \frac{ (\mxs-\mys)(\mys-\mzs) }{2\mys}.
\end{equation}
This corresponds to the familiar massless collinear limit.

The endpoint is the greater of eq.~(\ref{33m123extremA}) and the appropriate choice of eq.~(\ref{33m123extremB1}) or eq.~(\ref{33m123extremB2}) for the given value of $\mx$.  The precise conditions for choosing the correct endpoint are given in table \ref{33table} and derived in detail in Appendix \ref{m123lowapp}.

\subsection{Ordered endpoints for triplets equivalent to $m_{134}^2$}
There are likewise two ordered endpoints for the indistinguishable triplets $m_{134}^2$ and $m_{234}^2$ which will be called $(m_{134}^2)_\high$ and $(m_{134}^2)_\low$.  The derivation of these endpoints is very similar to what we have already seen for $m_{123}^2$.  The results are summarized here and derived in detail in the appendices.

\subsubsection{High endpoint}\label{m134hightxt}
Our calculations have all been performed in the $\partY$ rest frame where the event is symmetric under the exchange $\{\partX,1,2\}\leftrightarrow\{\partZ,3,4\}$.  Therefore the calculation of $(m_{134}^2)_\high$ proceeds in an entirely similar way and we can confirm that the two solutions are the same as for $\max(m_{123}^2)_\high$ with the opposite mass hierarchy requirements corresponding to the opposite ordering.  Details are given in Appendix \ref{m134highapp} and the results are listed in table \ref{33table}.


\subsubsection{Low endpoint}\label{m134lowtxt}
We follow the same technique as for $(m_{123}^2)_\low$ and find an expression for $m_{134}^2$ in terms of the composite masses $\mas$ and $\mbs$.  The result is shown in Appendix \ref{m134lowapp}, eq.~(\ref{33m134lowendpointexpn}).
This has two extrema on the boundary of the physical region in $(\ma,\mb)$, one of which is always the maximum, located along the edge $\mas=0$ at a value of $\mbs$ given in eq.~(\ref{33m134lowmbexpn}) where 
\begin{equation}
	{
	m_{234}^2 = \left(\sqrt{\frac{\mxs+\mys}{2}}-\sqz\right)^2 .
	}
\end{equation}
There are no extrema in the interior.
If
\begin{equation}
	\mxmys > \frac{2\mys-\mzs}{\mzs} .
\end{equation}
the implied value of $\mbs$ in eq.~(\ref{33m134lowmbexpn}) becomes negative and instead the maximum saturates to the corner $(\ma,\mb)=(0,0)$ where
\begin{equation}
	m_{234}^2 = \frac{ (\mxs-\mys)(\mys-\mzs) }{2\mys}.
\end{equation}
It is shown in Appendix \ref{m134lowapp} that there are no other maxima.

\begin{table}\renewcommand{\arraystretch}{2}
\caption{Table of endpoints for the 3+3 decay chain}\label{33table}
\begin{tabular}{| c | c c |}
\hline
$m_{12}^2$ & $(\mx - \my)^2$ &\\ \hline
$m_{34}^2$ & $(\my - \mz)^2$  &\\ \hline
$m_{1234}^2$ & $(\mx - \mz)^2$ &\\ \hline
$(m_{13}^2)_\highest$ & $\frac{(\mxs-\mys)(\mys-\mzs)}{\mys}$ &\\ \hline
$(m_{13}^2)_\high$ & $\frac{(\mxs-\mys)(\mys-\mzs)}{2\mys}$ &\\ \hline
$(m_{13}^2)_\low$ & numerically maximize eq.~(\ref{m23loweqn}) &\\ \hline
$(m_{13}^2)_\lowest$ & $\frac{(\mxs-\mys)(\mys-\mzs)}{4\mys}$ &\\ \hline

\multirow{2}{*}{$(m_{123}^2)_\high$} & $\frac{(\mxs-\mys)(\mys-\mzs)}{\mys}$ & $\mymz>\mxmy$ \\ 
& $(\mx-\mz)^2$ & otherwise \\ \hline

\multirow{3}{*}{$(m_{123}^2)_\low$} & $\frac{(\mxs-\mys)(\mys-\mzs)}{2\mys}$ & 
$\left\{ \begin{array}{l}
	\mymzs>{7+4\sqrt 3}\ \mathrm{ and }\  \mxmys<\frac{2\mys}{\mys+\mzs} \\
	\mymzs<{7+4\sqrt 3}\ \mathrm{ and }\ \mxmy < \frac{\mys-\mzs+2\my\mz}{\mys+\mzs}
	\end{array}\right.$ \\ 
& $\left( \mx-\sqrt{ \frac{\mys+\mzs}{2} } \right)^2$ & \makecell{$\mymzs>{7+4\sqrt 3}$ and\\ $\frac{2\mys}{\mys+\mzs} < \mxmys < \frac{1}{4\mys} \left( \sqrt{2(\mys+\mzs)} +\my+\mz\right)^2$} \\
& $(\mx-\my)(\mx-\mz)$ & otherwise \\ \hline

\multirow{2}{*}{$(m_{134}^2)_\high$} & $\frac{(\mxs-\mys)(\mys-\mzs)}{\mys}$ & $\mxmy>\mymz$ \\ 
& $(\mx-\mz)^2$ & otherwise \\ \hline

\multirow{2}{*}{$(m_{134}^2)_\low$} & $\frac{(\mxs-\mys)(\mys-\mzs)}{2\mys}$  & $\mxmys > \frac{2\mys-\mzs}{\mzs}$ \\ 
& $ \left( \sqrt{ \frac{\mxs+\mys}{2} }- \mz \right)^2 $& otherwise\\ \hline

$m_{1234}^2$ & $(\mx-\mz)^2$ &\\

\hline
\end{tabular}
\end{table}

\section{Decay chains containing a single 3-body vertex}\label{223sec}

We now turn to the computation of the kinematic endpoints of decay chains containing a single 3-body decay.  These decay chains are depicted in figure \ref{decaysfig} and are labelled according to our convention as 2+2+3, 2+3+2, and 3+2+2, depending on the position of the 3-body vertex in the decay chain.  We label the massive particles in this case $W$ through $Z$.  The results will be collected in tables \ref{223table} -- \ref{232table}.

Many of the endpoints can be taken directly from results for shorter decay chains by ignoring the parts of the chain before or after the relevant particles.  For endpoints such as $m_{123}^2$ in the 2+2+3 decay, results can also be copied from a similar decay without particle 4, since as described in \S\ref{33m13} and Appendix \ref{m23highestapp}, the maximum for an endpoint like this will be obtained in the limit that particle 4 is soft.  For these simple cases, results are listed in the tables but are not rederived here.  Below we discuss cases warranting further attention.

\begin{table}\renewcommand{\arraystretch}{2}
\caption{Table of endpoints for the 2+2+3 decay chain}\label{223table}
\begin{tabular}{| c | c c |}
\hline

$m_{12}^2$ & $\frac{(\mws-\mxs)(\mxs-\mys)}{\mxs}$ & \\ \hline
$m_{34}^2$ & $(\my-\mz)^2$ &\\ \hline
$(m_{23}^2)_\high$ & $\frac{(\mxs-\mys)(\mys-\mzs)}{\mys}$ & \\ \hline
$(m_{23}^2)_\low$ & $\frac{(\mxs-\mys)(\mys-\mzs)}{2\mys}$ & \\ \hline
$(m_{13}^2)_\high$ & $\frac{(\mws-\mxs)(\mys-\mzs)}{\mys}$ & \\ \hline
$(m_{13}^2)_\low$ & $\frac{(\mws-\mxs)(\mys-\mzs)}{2\mys}$ & \\ \hline

\multirow{4}{*}{$(m_{123}^2)_\high$} &  $\frac{(\mws-\mxs)(\mxs-\mzs)}{\mxs}$ & $\mwmx>\mxmy\mymz$ \\
& $\frac{(\mws\mys - \mxs\mzs)(\mxs-\mys)}{\mxs\mys}$ & $\mxmy > \mymz\mwmx$ \\
& $\frac{(\mws-\mys)(\mys-\mzs)}{\mys}$ & $ \mymz > \mwmx\mxmy $ \\
& $(\mw-\mz)^2$ & otherwise \\ \hline

$(m_{123}^2)_\low$ & \multicolumn{2}{l|}{Maximize over edges as described in Appendix \ref{223m123low}}  \\ \hline

\multirow{2}{*}{$m_{134}^2$} & $\frac{(\mws-\mxs)(\mys-\mzs)}{\mys}$  & $\frac{\sqrt{\mws-\mxs+\mys}}{\my} > \mymz$ \\
& $\left(\sqrt{\mws-\mxs+\mys}-\mz\right)^2$ & otherwise \\ \hline

\multirow{2}{*}{$m_{234}^2$} & $\frac{(\mxs-\mys)(\mys-\mzs)}{\mys}$  & $\mxmy>\mymz$ \\
& $(\mx-\mz)^2$ & otherwise \\ \hline

\multirow{3}{*}{$m_{1234}^2$}  &  $\frac{(\mws-\mxs)(\mxs-\mzs)}{\mxs}$ & $\mwmx>\mxmy\mymz$ \\
& $\frac{(\mws\mys - \mxs\mzs)(\mxs-\mys)}{\mxs\mys}$ & $\mxmy > \mymz\mwmx$ \\
& $(\mw-\mz)^2$ & otherwise \\

\hline
\end{tabular}
\end{table}

\subsection{Conversion to nearest neighbor endpoints}
Endpoints of invariant masses constructed from particles which are not all neighbors in the decay chain can be obtained from similar endpoints where all particles are neighbors \cite{Gjelsten:2005aw}. As an example, consider the endpoint of $m_{134}^2$ in the 2+2+3 decay, viewed in the $X$ frame.  As usual, we will want particle 1 and $B=3+4$ to be back to back.  The magnitude of the momentum of particle 2 is fixed by the masses of $X$ and $Y$.  The momentum of $B$ as measured in the $X$ frame will be boosted by the velocity of $Y$.  The endpoint must occur when this boost is as great as possible, which implies that $Y$ should be aligned with $B$ and particle 2 should be aligned with particle 1. All momenta are now fixed as functions of the masses.  
Consider the momentum of particle 1 as view in the $Y$ frame.  By solving for the mass of a hypothetical particle $V$ that would give this momentum to a massless particle when decaying to that particle and particle $Y$ in the $Y$ frame, we have transformed the problem into an endpoint involving only neighboring particles in the decay $V\to1Y\to 134Z$.

It is straightforward to show that the appropriate mass for $V$ in the 2+2+3 case is $m_V^2 = \mws-\mxs+\mys$, so that by making the replacement $\mx\to m_V$ we obtain the $m_{13}^2$ and $m_{134}^2$ endpoints from the results for $m_{23}^2$ and $m_{234}^2$, respectively.  

Similarly, the endpoints for $m_{14}^2$ and $m_{124}^2$ in the 3+2+2 decay can be obtained from the results for $m_{13}^2$ and $m_{123}^2$ by substituting $\my\to m_V=\mz\mx/\my$.


\subsection{Endpoints for quadruplet $m_{1234}^2$}\label{223m1234txt}
In calculating the endpoint for $m_{1234}^2$ in the 2+2+3 decay, we will depart from our usual convention by labeling the combination of particles 2, 3, and 4 as $B=2+3+4$.
The calculation then proceeds as in the case of $m_{134}^2$ for the 3+3 decay chain as described in \S\ref{m134hightxt}, except that now, in addition to ensuring that $\mbs>0$, we must also check that $\mbs<\max(\mbs)$ where $\max(\mbs)$ is the endpoint $m_{234}^2$ given in table \ref{223table}. This leads to three possible forms for the endpoint, depending on the mass spectrum.  The maximum possible invariant mass $\mw-\mz$ is again achieved when $B$ is able to balance the momentum of particle 1 so that $W$ and $Z$ are at rest in the same frame.  For some mass spectra, this is impossible, and $B$ collapses to the massless limit giving the familiar result for that case.  Meanwhile for other spectra, $B$ reaches its maximum value and the endpoint saturates to the form 
\begin{equation}
	m_{1234}^2 = \frac{(\mws\mys - \mxs\mzs)(\mxs-\mys)}{\mxs\mys} .
\end{equation}
The conditions for these different expressions to be the maximum are given in table \ref{223table} and derived in detail in Appendix \ref{223m1234}.

The results for the 3+2+2 are obtained in an analogous way and are related to the 2+2+3 results by symmetry.

The 2+3+2 decay also shows broadly similar results, but with an even higher degree of symmetry.  Again, for some mass spectra it is possible to balance the momenta and achieve the maximum possible invariant mass.  However, if the mass ratio $\my/\mz$ is large, the endpoint is obtained when particles 2 and 3 collapse and become collinear with particle 1, so that the three of them are effectively a single massless particle recoiling against particle 4, which receives a large amount of energy from the $Y$ decay to the much lighter $Z$. The decay can be viewed as $W\to 123Y\to 1234Z$.  It is easy to see that the endpoint listed in table \ref{232table} would be obtained.  Likewise, if $\mw/\mx$ is large, particles 2 and 3 collapse in the other direction and the decay is equivalent to $W\to 1X\to 1234Z$ with the corresponding result.  

\begin{table}\renewcommand{\arraystretch}{2}
\caption{Table of endpoints for the 3+2+2 decay chain}\label{322table}
\begin{tabular}{| c | c c |}
\hline

$m_{12}^2$ & $(\mw-\mx)^2$ & \\ \hline
$m_{34}^2$ & $\frac{(\mxs-\mys)(\mys-\mzs)}{\mys}$  &\\ \hline
$(m_{23}^2)_\high$ & $\frac{(\mws-\mxs)(\mxs-\mys)}{\mxs}$ & \\ \hline
$(m_{23}^2)_\low$ & $\frac{(\mws-\mxs)(\mxs-\mys)}{2\mxs}$ & \\ \hline
$(m_{24}^2)_\high$ & $\frac{(\mws-\mxs)(\mys-\mzs)}{\mys}$ & \\ \hline
$(m_{24}^2)_\low$ & $\frac{(\mws-\mxs)(\mys-\mzs)}{2\mys}$ & \\ \hline

\multirow{2}{*}{$m_{123}^2$} & $\frac{(\mws-\mxs)(\mxs-\mys)}{\mxs}$  & $\mxmy>\mwmx$ \\
& $(\mw-\my)^2$ & otherwise \\ \hline

\multirow{2}{*}{$m_{124}^2$} & $\frac{(\mws-\mxs)(\mys-\mzs)}{\mys}$  & $\mymz>\mwmx$ \\
& $\left(\mw-\frac{\mz\mx}{\my}\right)^2$ & otherwise \\ \hline

\multirow{4}{*}{$(m_{134}^2)_\high$} &  $\frac{(\mws-\mxs)(\mxs-\mzs)}{\mxs}$ & $\mwmx>\mxmy\mymz$ \\
& $\frac{(\mws\mys - \mxs\mzs)(\mxs-\mys)}{\mxs\mys}$ & $\mxmy > \mymz\mwmx$ \\
& $\frac{(\mws-\mys)(\mys-\mzs)}{\mys}$ & $ \mymz > \mwmx\mxmy $ \\
& $(\mw-\mz)^2$ & otherwise \\ \hline

\multirow{4}{*}{$(m_{134}^2)_\low$} & $\frac{(\mws-\mxs)(\mxs-\mzs)}{2\mxs}$ & $\mwmx > \frac{2\mxs}{\mzs}-1 $ \\ 
& $\frac{(\mws+\mxs-2\mys)(\mys-\mzs)}{2\mys}$ & $\mxmy<\mymz$ and $\mwmxs<\frac{2\my^4}{\mxs\mzs} -1$ \\
& $\frac{(\mxs-\mys)(\mws\mys + \mxs\mys - 2\mxs\mzs)}{2\mxs\mys}$ & $\mxmy>\mymz$ and $\mwmxs<\frac{2\mxs\mzs}{\my^4} -1$\\
& $\left(\sqrt{\frac{\mws+\mxs}{2}} - \mz\right)^2 $& otherwise \\ \hline

\multirow{3}{*}{$m_{1234}^2$}  &  $\frac{(\mws-\mys)(\mys-\mzs)}{\mys}$ & $\mymz > \mxmy\mwmx$ \\
& $\frac{(\mws\mys - \mxs\mzs)(\mxs-\mys)}{\mxs\mys}$ & $\mxmy > \mymz\mwmx$ \\
& $(\mw-\mz)^2$ & otherwise \\

\hline
\end{tabular}
\end{table}

\subsection{Low endpoints for triplet invariant masses}

\subsubsection{Endpoint for $(m_{123}^2)_\low$ in the 2+2+3 decay chain}\label{223m123lowtxt}
 This endpoint can be treated in the same way as the corresponding endpoint in the 3+3 decay chain.  The only difference is that the mass of the $A=1+2$ system now cannot exceed $\max(\mas)=\max(m^2_{12})=(\mws-\mxs)(\mxs-\mys)/\mxs$.  
In \S\ref{m123lowtxt} and Appendix \ref{m123lowapp} we saw that there is only a minimum in the interior of the physical region. The maximum must be on the edges.  A  variety of results are obtained for various mass spectra.  These are described in Appendix \ref{223m123low}.
In some regions of parameter space it is easy to write down an analytical expression for the endpoint, but in practice, it is easier to simply numerically maximize over the edges of the physical region, rather than exhaustively list all possibilities.

\subsubsection{Endpoint for $(m_{134}^2)_\low$ in the 3+2+2 decay chain}\label{322m234lowtxt}
The calculation of $(m_{134})_\low$ is again very similar to the 3+3 case discussed in \S\ref{m134lowtxt}, where there were only two possibilities, depending on whether or not the maximum on the $\ma=0$ edge saturated at $\mb=0$.  
The greatest possible value of $\mb$ was $\my-\mz$, and the location of the maximum never exceeded that bound.  Now, however, the greatest possible value of $\mbs=(\mxs-\mys)(\mys-\mzs)/\mys$, and it is possible for the implied location of the maximum to go beyond this point.
 This can happen under two different conditions, but the results are simple and are listed in table \ref{322table}.  The derivation can be found in Appendix \ref{322m234low}.

\begin{figure}
\includegraphics[scale=0.75]{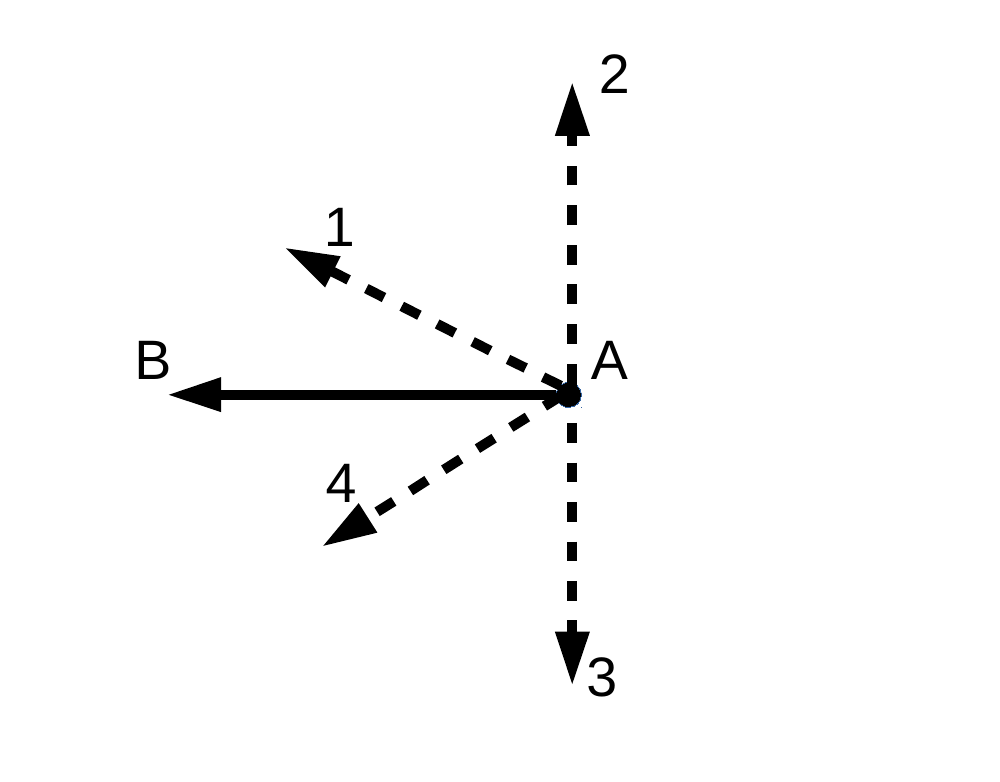}
\caption{The symmetrical configuration in the rest frame of $A=2+3$ required to obtained the $(m_{124}^2)_\low$ endpoint in the 2+3+2 decay chain.}\label{232fig}
\end{figure}

\subsubsection{Endpoint for $(m_{124}^2)_\low$ in the 2+3+2 decay chain}\label{232m124lowtxt}
In the previous calculations of low triplet endpoints, it was convenient to work in the frame of the intermediate particle which followed the first two visible particles and preceded the last two.  This frame possesses a kind of symmetry in which the first two and last two visible particles could be paired together as an effective massive particle. The beginning and end of the decay could then be treated as two independent 2-body decays.  Here such a simple partition is not possible since the middle two particles come from the same vertex.  The most symmetrical choice will be to work in the rest frame of the (2,~3) system, which we will in this section denote $A$, as shown in the lower left panel of figure \ref{decaysfig}.  We will also denote the (1,~4) system as $B$.  As argued previously, the low endpoint occurs when $m_{124}^2=m_{134}^2$.  In order to enforce this equality in the $A$ rest frame, where particles 2 and 3 are back to back, we must align $B$ perpendicularly to 2 and 3, as depicted in figure \ref{232fig}.  Finally, grouping particles $W$ and $Z$ as $C=W-Z$, this can be viewed as a 2-body decay $C\to AB$. Note that $m_C^2 = (p_W^\mu - p_Z^\mu)^2 = (p_1^\mu+ p_2^\mu+ p_3^\mu+ p_4^\mu)^2 = m_{1234}^2$. In the $A$ frame the invariant mass of this configuration is simply
\begin{equation}
	m_{124}^2 = m_{2B}^2 = m_B^2 + 2p_2 E_B .
\end{equation}
Clearly $p_2 = m_A/2$ and $E_B = (m_C^2 - m_A^2 - m_B^2)/2m_A$.  We see that
\begin{equation}
	m_{124}^2 =\frac{1}{2}\left( m_C^2 + m_B^2 - m_A^2 \right).
\end{equation}
This is a monotonic function of all three masses, so the maximum must lie on the boundary of the physical region.  Maximization is best handled numerically.  A prescription for doing this is given in  Appendix \ref{232m124low}.

\begin{table}\renewcommand{\arraystretch}{2}
\caption{Table of endpoints for the 2+3+2 decay chain}\label{232table}
\begin{tabular}{| c | c c |}
\hline

$(m_{12}^2)_\high$ & $\frac{(\mws-\mxs)(\mxs-\mys)}{\mxs}$  & \\ \hline
$(m_{12}^2)_\low$ & $\frac{(\mws-\mxs)(\mxs-\mys)}{2\mxs}$  & \\ \hline
$(m_{34}^2)_\high$ & $\frac{(\mxs-\mys)(\mys-\mzs)}{\mys}$  &\\ \hline
$(m_{34}^2)_\low$ & $\frac{(\mxs-\mys)(\mys-\mzs)}{2\mys}$  &\\ \hline
$m_{23}^2$ & $(\mx-\my)^2$ & \\ \hline
$m_{14}^2$ & $\frac{(\mws-\mxs)(\mys-\mzs)}{\mys}$  & \\ \hline

\multirow{2}{*}{$m_{123}^2$} & $\frac{(\mws-\mxs)(\mxs-\mys)}{\mxs}$  & $\mwmx>\mxmy$ \\
& $(\mw-\my)^2$ & otherwise \\ \hline

\multirow{4}{*}{$(m_{124}^2)_\high$} &  $\frac{(\mws-\mxs)(\mxs-\mzs)}{\mxs}$ & $\mwmx>\mxmy\mymz$ \\
& $\frac{(\mws\mys - \mxs\mzs)(\mxs-\mys)}{\mxs\mys}$ & $\mxmy > \mymz\mwmx$ \\
& $\frac{(\mws-\mys)(\mys-\mzs)}{\mys}$ & $ \mymz > \mwmx\mxmy $ \\
& $(\mw-\mz)^2$ & otherwise \\ \hline

$(m_{124}^2)_\low$ & \multicolumn{2}{l|}{Numerically maximize eq.~(\ref{232m124maximize}) subject to eq.~(\ref{232m124bc})}\\ \hline

\multirow{2}{*}{$m_{234}^2$} & $\frac{(\mxs-\mys)(\mys-\mzs)}{\mys}$  & $\mymz>\mxmy$ \\
& $(\mx-\mz)^2$ & otherwise \\ \hline

\multirow{3}{*}{$m_{1234}^2$}  &  $\frac{(\mws-\mys)(\mys-\mzs)}{\mys}$ & $\mymz > \mxmy\mwmx$ \\
& $\frac{(\mws-\mxs)(\mxs-\mzs)}{\mxs}$ & $\mwmx > \mymz\mxmy$ \\
& $(\mw-\mz)^2$ & otherwise \\

\hline
\end{tabular}
\end{table}

\section{Conclusions}\label{conclusion}

We have presented kinematic endpoints for 5-body decays consisting of a cascade of 2- and 3-body decays.  In cases where a given invariant mass could be formed in more than one indistinguishable way   using particles from a common vertex, we have presented techniques for calculating the ordered endpoints. 
We give analytical expressions for all leading endpoints.  Where possible, we exploit considerations of symmetry to find analytical expressions for the sub-leading ordered endpoints as well.  In some cases, however, the final step of the calculation must be done numerically.  In these cases, we have given a simple parameterization of the remaining phase space degrees of freedom in terms of the masses of pseudoparticles.

These kinematic endpoints are relevant to the determination of the masses of invisible particles involved in scenarios such as gluino decay to a pair of quarks, a pair of leptons, and an LSP, as well as other scenarios with similar final states.  In order to solve for the unknown masses, at least one linearly independent endpoint is needed for each. Not all endpoints are linearly independent.  For example, the results for the 3+2+2 decay chain in table \ref{322table} show that when $\mx/\my>\my/\mz$, the endpoints for $m_{23}^2$ and $m_{123}^2$ are identical.  The high and low endpoints for $m_{23}^2$ are also identical up to a factor of 2.  This would also be true for a 4-body decay equivalent to the first two stages of the 3+2+2 decay.  
However, the ordered endpoints for invariant masses that span the full length of the 5-body decay chain, such as $m_{123}^2$ in table \ref{33table} or $m_{134}^2$ in table \ref{322table},  have high and low endpoints that are linearly independent in general.  
We see that in high multiplicity decays, even when some of the final state particles are kinematically indistinguishable, the ordered endpoints contain valuable information about the mass spectrum.  These endpoints must be considered in order to take full advantage of the kinematic information available in the decay.
 
\acknowledgments
The author gratefully acknowledges valuable comments and guidance from Can Kilic. 
This work is supported by the National Science Foundation under Grant Number PHY-1620610.

\bibliography{endpointbib}{}
\bibliographystyle{utphys}

\begin{appendix}

\section{Appendix: $(m_{13}^2)_\highest$ in the 3+3 decay chain}\label{m23highestapp}
We can verify the result of \S(\ref{m23highesttxt}) by writing the exact expression $m_{13}^2=2p_1 p_3 (1-\cos\theta)$ where $\theta$ is the angle between the 3-momenta of 1 and 3, and $p_i$ is the magnitude of the 3-momentum of particle $i$.  For any values of the momenta, this will be maximized when 1 and 3 are back-to-back, that is, when $\cos\theta=-1$. For convenience, we can define an initial state pseudoparticle $\tilde \partX=\partX-2$ and a final state pseudoparticle $\tilde \partZ=4+\partZ$.  In terms of these, each vertex now represents a 2-body decay and the magnitudes of the 3-momenta of particles 1 and 3 are fixed by the masses of the other particles to be $p_1^2=[0,Y,\tilde X]$ and $p_3^2=[0,Y,\tilde Z]$ .
The pseudoparticle masses are $\mxtils = \mxs-2p_2(E_X-p_X\cos\theta_{2X})$ and $ \mztils = \mzs+2p_4(E_Z-p_Z\cos\theta_{4Z})$, where the angles are measured between the two particles in the subscript.
Using these results, we obtain
\begin{equation}
	m_{23}^2 = \frac{( \mxtils-\mys)(\mys-\mztils)}{\mys}.
\end{equation}
This expression is monotonic in $\mxtil$ and $\mztil$.  To maximize it, we should choose the highest possible value for $ \mxtil$ and the lowest possible value for $ \mztil$.  These are $\max( \mxtil)=\mx$ and $\min( \mztil)=\mz$, achieved when $p_1=p_4=0$.  Making these substitutions, we obtain the result eq.~(\ref{33m13highestresult}).

\section{Appendix: $(m_{13}^2)_\low$ in the 3+3 decay chain}\label{m23lowapp}
This endpoint occurs at the maximum value of the invariant mass obtained when the system of equations eq.~(\ref{firstloweq})--(\ref{lastloweq}) in \S\ref{m23lowtxt} is satisfied.  The invariant mass is 
\begin{multline}\label{m23loweqn}
	m_{13}^2(b,c) =  
		\frac{\sqrt{c^2 (\mys-\mzs)+\mzs}-\my}{2 (b-1) \left(c^2-1\right) (b+c)^2} \times \\
		\times \Bigg(\my \left(b^2\left(2 c^2+ c+2 \right)+b\left( c^3+2  c^2+4  c\right)+c^3+2 c^2+b^3 c\right)
		 +(b+c)^2 s\my 
		  -\frac{1}{2} \left(s+b c+1\right)\times	\\  
		 \times \sqrt{4 \mys \left(-s^2-(c+1) s+b c (b c+c+3)+c+2\right)^2-8 (b-1) (c+1) (b+c)^2 \left(-s+b c+1\right) (\mxs-\mys)}\Bigg) , 
\end{multline}
where
\begin{equation}
	s\equiv \sqrt{(1-b^2)(1-c^2)} = \sin\theta \sin\phi.
\end{equation}
This can be maximized numerically over the domain $b,c\in [0,1]$.

\section{Appendix: $(m_{123}^2)_\low$ in the 3+3 decay chain}\label{m123lowapp}
Here we detail the calculation summarized in \S(\ref{m123lowtxt}).
Since we are looking for configurations which satisfy $m_{123}^2=m_{124}^2$, we demand that the magnitudes of the momenta $p_3$ and $p_4$ are equal in the $Y$ frame, and that each makes the same angle with respect to the direction of the momentum of the $A=1+2$ system.
Let $p=p_3=p_4$. 
This can be related to the momentum of the composite particle $\partB=3+4$ and the kinematic function eq.~(\ref{kinematicfn}) by
\begin{equation}
	2 p \cos \theta = [B,Y,Z]^{1/2} ,
\end{equation}
where $\theta$ is the angle between particle 3 or 4 and the direction anti-parallel to the momentum of the $A$ system in the $Y$ frame. 
The energy of $\partB$ is just $2p$ so its squared mass is
\begin{equation}
	\mbs = (2p)^2 - [B,Y,Z].
\end{equation}
Alternatively, the squared mass can be computed from the Lorentz product of the 4-momenta of particles 3 and 4 as
\begin{equation}
	\mbs = 2p^2(1-\cos 2\theta).
\end{equation}
This set of equations can be solved to yield
\begin{equation}
	p=\frac{\mys+\mbs-\mzs}{4\my}
\end{equation}
\begin{equation}
	\cos 2\theta = 1-\frac{8 \mbs \mys}{(\mbs+\mys-\mzs)^2}  .
\end{equation}
Employing these results, we find $m_{123}^2$ as a function of the composite particle masses $\mas$ and $\mbs$,
\begin{equation}\label{33m123lowendpointexpn}
	m_{123}^2 = \mas + \frac{\mys+\mbs-\mzs}{2\my}
		\left( \frac{\mxs-\mys-\mas}{2\my} + \sqrt{[A,Y,X]}\sqrt{1-
			\frac{4\mbs\mys}{(\mys+\mbs-\mzs)^2}}
		\right) .
\end{equation}
We will explore the structure of the extrema of this expression as a function of $\mx$.  

There are a pair of extrema on the boundaries of the physical region defined by $\ma\in [0,\sqx-\sqy]$, $\mb\in [0,\sqy-\sqz]$.  In the corner $(\ma,\mb)=(\sqx-\sqy,\ \sqy-\sqz)$ we obtain the value 
\begin{equation}\label{m123_iii}
	m_{123}^2=(\sqx-\sqy)(\sqx-\sqz). 
\end{equation}
Along the $\mb=0$ edge at 
\begin{equation}\label{33m123edgemavalue}
	\ma = \left(\mxs+\mys-\frac{\mx(3\mys+\mzs)}{\sqrt{2(\mys+\mzs)}}\right)^{1/2}
\end{equation}
there is an extremum with value
\begin{equation}\label{m123_ii}
	m_{123}^2=\left(\mx - \sqrt{\frac{\mys+\mzs}{2}}\right)^2 . 
\end{equation}

There is also an extremum off the boundary at $(\ma,\mb) = \left(\mz, \sqrt{2(\mys+\mzs)-\mxs}\right)$ with value
\begin{equation}\label{m123_i}
	m_{123}^2=\frac{\mxs-\mys-\mzs}{2} .
\end{equation} 
 However, this extremum does not always exist in the physical region.  The requirement $\ma < \sqx-\sqy$ (or equivalently $\mb < \sqy-\sqz$) implies that this extremum disappears when $\mx < \sqy+\sqz$.  Meanwhile, the requirement $\mb>0$ excludes the region $\mx>\sqrt{2(\mys+\mzs)}$.  We will show that this extremum, when it does exist, is a minimum. 

At the limits of validity of eq.~(\ref{m123_i}), it coincides with the two boundary extrema.  That is, when $\mxs = (\sqy+\sqz)^2$, eq.~(\ref{m123_i}) and eq.~(\ref{m123_iii}) are equal, and when $\mxs=2(\mys+\mzs)$, eq.~(\ref{m123_i}) and eq.~(\ref{m123_ii}) are equal.  Their derivatives with respect to $\mxs$
\begin{equation}
	\frac{d}{d\mxs}(\mathrm{eq.} (\ref{m123_iii})) =
	1- \frac{\sqy+\sqz}{2\sqx}
\end{equation}
\begin{equation}
	\frac{d}{d\mxs}(\mathrm{eq.} (\ref{m123_ii})) =
	1- \sqrt\frac{\mys+\mzs}{2\mxs}
\end{equation}
 are also equal at these points.  However, the derivatives are monotonically increasing with $\mxs$ while the derivative of eq.~(\ref{m123_i}) is constant, so that eq.~(\ref{m123_ii}) and eq.~(\ref{m123_iii}) always exceed eq.~(\ref{m123_i}), which is therefore a minimum when it exists.  

Eq.~(\ref{m123_ii}) and eq.~(\ref{m123_iii}) intersect at 
\begin{equation}
	(\mxs)_\mathrm{eq} = \frac{1}{4} \left(\sqrt{2 (\mys+\mzs)}+\my+\mz \right)^2 ,
\end{equation}
halfway in $\mxs$ between the limits of existence of the minimum.
When $\mxs<(\mxs)_\mathrm{eq}$, eq.~(\ref{m123_ii}) is the global maximum in the physical region.  When $\mxs>(\mxs)_\mathrm{eq}$, eq.~(\ref{m123_iii}) is the maximum.

However, the preceding discussion assumes that the location corresponding to eq.~(\ref{m123_ii}) is indeed on the boundary of the physical region, that is, that eq.~(\ref{33m123edgemavalue})~$>0$.
This will be true when 
\begin{equation}\label{m123_corner_extremum_condition}
	\frac{\mxs}{\mys} > \frac{2\mys}{\mys+\mzs}.
\end{equation}
If this inequality fails, then the extremum hits the corner $(\ma,\mb)=(0,0)$ where
\begin{equation}\label{m123_corner_extremum}
	m_{123}^2 = \frac{(\mxs-\mys)(\mys-\mzs)}{2\mys} .
\end{equation}
This will still be greater than eq.~(\ref{m123_iii}) as long as $\mys/\mzs > 7+4\sqrt3$.
Indeed, if this inequality is violated, $2\my^4/(\mys+\mzs) > (\mxs)_\mathrm{eq}$ so that eq.~(\ref{m123_corner_extremum_condition}) implies that eq.~(\ref{m123_ii}) is never a maximum at all.
If this is the case, eq.~(\ref{m123_iii}) is the maximum until 
\begin{equation}
	\mxmy < \frac{\mys-\mzs+2{\my\mz}}{\mys+\mzs} ,
\end{equation}
when eq.~(\ref{m123_corner_extremum}) finally overtakes eq.~(\ref{m123_iii}).

\section{Appendix: $(m_{134}^2)_\high$ in the 3+3 decay chain}\label{m134highapp}
As noted in \S\ref{m134hightxt}, this endpoint has a simple relationship to $(m_{123}^2)_\high$ and the calculation is very similar to that presented in \S\ref{m123hightxt}.  We begin by combining particles 2 and $\partX$ into $\tilde\partX = X-2$.  It is immediately clear that $m_{134}^2$ will be maximized when particle 2 is soft and $ \mxtil=\mx$.  In terms of the composite particle $B=3+4$ we obtain
\begin{equation}
	m_{134}^2 = \mbs+\frac{\mxs-\mys}{2\mys}\left(\mys+\mbs-\mzs+\sqrt{(\mys+\mzs-\mbs)^2-4\mys\mzs}\right),
\end{equation}
which has a maximum at 
\begin{equation}
	\mbs=\mys+\mzs - \frac{\mz}{\mx}(\mxs+\mys)
\end{equation}
corresponding to
\begin{equation}
	m_{134}^2 = (\sqx-\sqz)^2.	
\end{equation}
This maximum is inside the physical region $\mb\in[0,\sqy-\sqz]$ when $\mx/\my \le \my/\mz$.  Note that this is complementary to the condition for the analogous result for $m_{123}^2$.  If this condition is not satisfied, then the maximum again occurs at the boundary $\mb=0$ where
\begin{equation}
	m_{134}^2 = \frac{(\mxs-\mys)(\mys-\mzs)}{\mys}.
\end{equation}

\section{Appendix: $(m_{134}^2)_\low$ in the 3+3 decay chain}\label{m134lowapp}
Using the same symmetry arguments as applied to $(m_{123}^2)_\low$, we proceed to derive the results summarized in \S\ref{m134lowtxt}.  
We take the momentum $p=p_1=p_2$ of particle 1 or 2 to be directed at an angle $\theta$ with respect to the direction anti-parallel to the $B=3+4$ system.  The momentum can be related to the kinematic function eq.~(\ref{kinematicfn}) by
\begin{equation}
	2 p \cos \theta = [A,Y,X]^{1/2} .
\end{equation}
The energy of $\partA$ is just $2p$ so its squared mass is
\begin{equation}
	\mas = (2p)^2 - [A,Y,X].
\end{equation}
Alternatively, the squared mass can be computed from the Lorentz product of the 4-momenta of particles 1 and 2 as
\begin{equation}
	\mas = 2p^2(1-\cos 2\theta).
\end{equation}
This set of equations can be solved to yield
\begin{equation}
	p=\frac{\mxs-\mas-\mys}{4\my}
\end{equation}
\begin{equation}
	\cos 2\theta = 1- \frac{8\mas\mys}{(\mxs-\mas-\mys)^2}  .
\end{equation}
Employing these results, we find $m_{134}^2$ as a function of the composite particle masses $\ma$ and $\mb$,
\begin{equation}\label{33m134lowendpointexpn}
	m_{134}^2 = m_{2B}^2 = \mbs + \frac{\mxs-\mas-\mys}{2\my}
		\left( \frac{\mys+\mbs-\mzs}{2\my} +
		\sqrt{[B,Y,Z]}
		\sqrt{1-\frac{4\mas\mys}{(\mxs-\mys-\mas)^2}}
		\right) .
\end{equation}
We will explore the structure of the extrema of this expression as a function of $\mx$.  

This expression possesses an extremum at $(\ma,\mb) = (\sqrt{2(\mxs+\mys)-\mzs}, \mx)$ but this is clearly outside the physical region since $\mb\le \my<\mx$.  The only relevant extrema, then, are on the boundary of the physical region.  
One is along the $\ma=0$ edge at
\begin{equation}\label{33m134lowmbexpn}
	\mb =  \left(-\frac{\sqrt{2} \mys\mz}{\sqrt{ \mxs+\mys}}-\frac{\mz\sqrt{ \mxs+\mys}}{\sqrt{2}}+\mys+\mzs\right)^{1/2}
\end{equation}
with value
\begin{equation}\label{m234_b}
	m_{134}^2=  \left(\sqrt{\frac{\mxs+\mys}{2}}- \mz \right)^2 .
\end{equation}
The other is at $(\ma,\mb)=(\sqx-\sqy,\ \sqy-\sqz)$ where we obtain the value 
\begin{equation}\label{m234_a}
	m_{134}^2=(\sqx-\sqz)(\sqy-\sqz). 
\end{equation}

Note that when $\mx=\my$, its lowest possible value, both eq.~(\ref{m234_a}) and eq.~(\ref{m234_b}) and their derivatives
\begin{equation}
	\frac{d}{d\mxs}(\mathrm{eq.} (\ref{m234_b})) =
	\frac{1}{2} - \frac{\mz}{\sqrt{2(\mxs+\mys)}}
\end{equation}
\begin{equation}
	\frac{d}{d\mxs}(\mathrm{eq.} (\ref{m234_a})) =
	\frac{ \my -  \mz}{2 \mx}
\end{equation}
are equal.  However it is easily seen that the second derivative of eq.~(\ref{m234_a}) is always negative while that of eq.~(\ref{m234_b}) is positive.  As a result, we conclude that eq.~(\ref{m234_b}) is the maximum.

The preceding assumes that eq.~(\ref{m234_b}) is in fact on the boundary of the physical region.  Requiring $\mb>0$ shows that this maximum disappears when $\mxs/\mys>(2\mys-\mzs)/\mzs$, and is replaced by the value
\begin{equation}\label{m234_alt}
	m_{134}^2 = \frac{(\mxs-\mys)(\mys-\mzs)}{2\mys}
\end{equation}
located at $(\ma,\mb)=(0,0)$. Eq.~(\ref{m234_alt}) is greater than eq.~(\ref{m234_a}) at the point that it branches off eq.~(\ref{m234_b}) and has constant positive slope in $\mxs$.  It is therefore also always greater than eq.~(\ref{m234_a}) in the region where it is valid.  We see that the true maximum is always either eq.~(\ref{m234_b}) or eq.~(\ref{m234_alt}), depending on the value of $\mx$ relative to $\my$ and $\mz$.  Eq.~(\ref{m234_a}) is always a minimum.

\section{Appendix: $m_{1234}^2$ in the 2+2+3 decay chain}\label{223m1234}
Here we detail the calculation summarized in \S\ref{223m1234txt}.
This calculation proceeds just as the calculation for $m_{134}^2$ in the 3+3 decay, except that particle $B$ is taken to be the combination $B=2+3+4$.  There is a value of $\mb$ for which the endpoint reaches the greatest possible value 
\begin{equation}\label{m1234truemax}
	m_{1234}^2= (\mw-\mz)^2.
\end{equation}
By reviewing our previous results in Appendix~\ref{m134highapp}, we see that the mass of $\mb$ where this maximum would occur takes unphysical values $\mbs<0$ when 
\begin{equation}
	\mwmx > \mxmy\mymz ,
\end{equation}
in which case the endpoint saturates to the familiar value
\begin{equation}
	m_{1234}^2 = \frac{(\mws-\mxs)(\mxs-\mzs)}{\mxs}.
\end{equation}
However, unlike the previous calculation, $\mb$ cannot always take values as high as $\mx-\mz$.  Indeed, it cannot exceed
\begin{equation}\label{lowermaxmb}
	\max(\mb^2) = \frac{(\mws-\mys)(\mys-\mzs)}{\mys}
\end{equation}
if
\begin{equation}\label{mblowmaxexists}
	\mxmy > \mymz .
\end{equation}
When this is the case, the implied value of $\mbs$ that would give eq.~(\ref{m1234truemax}) will exceed eq.~(\ref{lowermaxmb}) if
\begin{equation}\label{mbexceedslowmax}
	\mxmy > \mymz\mwmx .
\end{equation}
However, $\mw/\mx>1$ so if eq.~(\ref{mbexceedslowmax}) holds, then so does eq.~(\ref{mblowmaxexists}).  In other words, if the value of $\mbs$ that gives eq.~(\ref{m1234truemax})  exceeds the value in eq.~(\ref{lowermaxmb}), then that value is indeed the maximum physically allowed value of $\mbs$ so that eq.~(\ref{m1234truemax}) cannot be achieved.  Instead, the endpoint saturates at this highest allowed value of $\mbs$, eq.~(\ref{lowermaxmb}), which gives the result listed in table \ref{223table}.

\section{Appendix: $(m_{123})_\low$ in the 2+2+3 decay chain}\label{223m123low}
As mentioned in \S\ref{223m123lowtxt}, this endpoint can be treated in the same way as the corresponding endpoint in the 3+3 decay chain.  The only difference is that the mass of the $A=1+2$ system now cannot exceed $\max(\mas)=\max(m^2_{12})=(\mws-\mxs)(\mxs-\mys)/\mxs$.  
Since we have seen that there is only a minimum in the interior of the region, we should look for the maximum along the edges defined by $\mb=0$ and $\ma=\max(\ma)$.

 The expression for the invariant mass is  
\begin{equation}\label{223m123endpointexpn}
m_{123}^2 = \mas+ \frac{\mbs+\mys-\mzs}{2\my} \left(\frac{\mws-\mas-\mys}{2\my}+\sqrt{ [A,Y,W]}\sqrt{1-\frac{4 \mbs \mys}{(\mbs+\mys-\mzs)^2}} \right).
\end{equation}
The $\max(\ma)$ extremum occurs at  
\begin{equation}\label{223m123mb}
\mbs = \mys+\mzs-\frac{\mz \left(\mws \mys+\mx^4-2 \mxs \mys\right)}{\mx\sqrt{ (\mws-\mxs) (\mxs-\mys)}} .
\end{equation}
The value of the invariant mass at the extremum can be found by evaluating eq.~(\ref{223m123endpointexpn}) at $\mas=\max(\mas)$ and the value of $\mbs$ in eq.~(\ref{223m123mb}).
We must also require that this value of $\mbs$ is physical.  This will be true if 
\begin{equation}
	\frac{\mzs}{\mys}\frac{\mxs-\mys}{\mys}+1 < \mwmxs < \frac{\mxs-\mys}{\mzs} +1 .
\end{equation}
Outside this range, the extremum will remain at $(\ma,\mb)=(\max(\ma),0)$.

On the $\mb=0$ edge, the extremum has the simple form
\begin{equation}
	m_{123}^2 = \left( \mw - \sqrt\frac{\mys+\mzs}{2} \right)^2 ,
\end{equation}
at a value of $\ma$
\begin{equation}
	\mas =\mws+\mys- \frac{\mw (3 \mys+\mzs)}{\sqrt{2 (\mys+\mzs)}},
\end{equation}
assuming this value is physical.
The location of this extremum will exceed $\max(\mas)$ outside the range
\begin{equation}
	\frac{\mxs}{\my^4} \frac{\mys+\mzs}{2} < \mwmxs < \frac{2\mxs}{\mys+\mzs} ,
\end{equation}
although the lower edge is sometimes less than one so that it is never achieved by $\mw/\mx$.  On the other hand, we find that the extremum dips below the physical bound $\ma=0$ when 
\begin{equation}
	\frac{\mws}{\mys} < \frac{2\mys}{\mys+\mzs} .
\end{equation}
Thus we see that for small values of $\mw$, the extremum will saturate to either the $\ma=\max(\ma)$ corner or the $\ma=0$ corner.
The condition for the latter to happen is 
\begin{equation}
	\mxmys < \frac{2\mys}{\mys+\mzs}.
\end{equation}

In some regions of parameter space, it is easy to see what the true maximum is.  For example, when the mass spectrum is such that both extrema on the boundary converge to the corner $(\ma,\mb)= (\max(\ma), 0)$, then the maximum is obviously given by eq.~(\ref{223m123endpointexpn}) evaluated at this point.  Similarly if the $\max(\ma)$ extremum hits the $\mb=0$ lower bound while the $\mb=0$ extremum also hits the $\ma=0$ lower bound, it is clear that the true maximum is at $(\ma,\mb)=(0,0)$ which has the value
\begin{equation}
	(m_{123}^2)_\low = \frac{(\mws-\mys)(\mys-\mzs)}{2\mys} .
\end{equation}
In practice, it is easier to simply numerically maximize eq.~(\ref{223m123endpointexpn}) over the two relevant edges of the physical region, rather than exhaustively list all possibilities.

\section{Appendix: $m_{134}^2$ in the 3+2+2 decay chain}\label{322m234low}
As mentioned in \S\ref{322m234lowtxt}, the calculation for this endpoint is very similar to the corresponding 3+3 case, except that we must check not only that the implied value of $\mb$ at the maximum does not go below zero but that it also does not exceed $\max(\mb^2) = \max(m_{34}^2) = (\mxs-\mys)(\mys-\mzs)/\mys$.  It will exceed this bound if
\begin{equation}
	\mwmx < \max\left\{
		\frac{2\my^4}{\mxs\mzs} -1,\ 
		\frac{2\mxs\mzs}{\my^4} -1
	\right\}.
\end{equation}
It is easy to verify that the first expression in brackets will be the relevant one when $\my/\mz>\mx/\my$.  If this occurs, then the endpoint saturates to one of the two expressions listed in table \ref{322table}, depending on the hierarchy between $\my/\mz$ and $\mx/\my$.  The other expressions for the endpoint are identical to the 3+3 case.

\section{Appendix: $m_{124}^2$ in the 2+3+2 decay chain}\label{232m124low}
As discussed in \S\ref{232m124lowtxt}, the endpoint is
\begin{equation}\label{232m124maximize}
	(m_{124}^2)_\low = \max\left[ \frac{1}{2}\left( m_{1234}^2+\mbs-\mas \right) \right],
\end{equation}
where the maximization is over all degrees of freedom in the decay.  This expression is linear in all squared masses, so the maximum will lie somewhere on the boundary of the physically allowed region.  
This decay can be viewed as a family of 4-body decays $W\to 1X\to 1AY\to 1A4Z$ distinguished by their value of $\ma$.  

For a 4-body decay, labeling the final state particles 1 through 4, the boundary of the physically allowed region of phase space is given by the condition \cite{Agrawal:2013uka,ByersYang} 
\begin{equation}
	\lambda(m_1^2, m_{12}^2, m_{123}^2, m_{1234}^2, m_{2}^2, m_{23}^2, m_{234}^2, m_{3}^2, m_{34}^2, m_{4}^2) = 0,
\end{equation}
where
\begin{equation}
	\lambda(m_1^2, m_{12}^2, m_{123}^2, m_{1234}^2, m_{2}^2, m_{23}^2, m_{234}^2, m_{3}^2, m_{34}^2, m_{4}^2) \equiv
	\left|
		\begin{array}{cccccc}
			0&m_{1}^2& m_{12}^2& m_{123}^2& m_{1234}^2& 1\\
			m_1^2& 0 & m_{2}^2& m_{23}^2 & m_{234}^2& 1\\
			m_{12}^2 & m_{2}^2 & 0 & m_{3}^2 & m_{34}^2 & 1\\
			m_{123}^2 & m_{23}^2 & m_{3}^2 & 0 & m_4^2 & 1\\
			m_{1234}^2 & m_{234}^2 & m_{34}^2 & m_4^2 & 0 & 1\\
			1&1&1&1&1&0\\
		\end{array}
	\right| .
\end{equation}

For any given value of $\ma$ and $\mb$, we can apply this boundary condition to our situation as
\begin{equation}\label{232m124bc}
	\lambda(0,\, m_{1A}^2,\, m_{1A4}^2,\, \mws,\, \mas,\, \mas+m_{1A4}^2-m_{1A}^2-\mbs,\, \mxs,\, 0,\, \mys,\, \mzs) =0.
\end{equation} 
This is a quadratic equation for all the squared masses and it can be easily solved for $m_{1A4}^2=m_{1234}^2$.  Substituting this solution into eq.~(\ref{232m124maximize}), we obtain an expression that can be maximized numerically over the remaining degrees of freedom, which are $m_{1A}^2$, $\mas$, and $\mbs$.

\end{appendix}
\end{document}